\documentclass[aps,prb,floatfix,reprint]{revtex4-1}
\usepackage{graphicx}
\usepackage{subfigure}
\usepackage{amsfonts}
\usepackage{amsmath}
\usepackage{amssymb}
\usepackage{subfigure}
\usepackage{soul}
\usepackage{xcolor,soul}
\usepackage[utf8]{inputenc}
\usepackage{amssymb,amsthm}
\usepackage{amsmath}
\usepackage{graphicx,color}
\usepackage{bm}

\newcommand{\im}[0]{\mathrm{i}}

\begin{document}

\title{Thermal Motors with Enhanced Performance due to Engineered Exceptional Points }

\author{Lucas J. Fern\'andez-Alc\'azar, Rodion Kononchuk, Tsampikos Kottos} 
\affiliation{Wave Transport in Complex Systems Lab, Department of Physics, Wesleyan University, Middletown, CT-06459, USA}
\date{\today}

\begin{abstract}
A thermal current, generated by a temperature gradient between two reservoirs coupled to a carefully designed photonic or (micro-)
electromechanical circuit, might induce non-conservative forces that impulse a mechanical degree of freedom to move along a {\it 
closed trajectory}. We show that in the limit of long - but finite - modulation periods, the extracted power and the efficiency of such 
autonomous motors can be maximized when an appropriately designed spatio-temporal symmetry violation is induced and when 
the motor operates in the vicinity of exceptional point (EP) degeneracies. These singularities appear in the spectrum of the effective 
non-Hermitian Hamiltonian that describes the combined circuit-reservoirs system when we judiciously tailor the coupling between 
them. In the photonic framework, these motors can be propelled by thermal radiation and can be utilized as autonomous self-powered 
microrobots, or as micro-pumps for microfluidics in biological environments. The same designs can be also implemented with 
electromechanical elements for harvesting ambient mechanical (or electrical) noise for powering a variety of auxiliary systems.
\end{abstract}

\maketitle

\section{Introduction} 

The manipulation of microscopic objects via currents has become an indispensable tool in many disciplines of 
science and technology, revolutionizing a variety of applications in areas as diverse as micro-engineering and 
micro-robotics, to biology and medicine \cite{CKOF16,WSCGLL16,KZDHF99,KBPMBF08,TMJBKMKS11,
KRPMKHEF11,PG13,CSGWSSBGPHM10,KSMLR14,WNMMMRB12}. Depending on the application, the source 
of these currents varies from thermal radiation and thermal vibrations to electrical and chemical energy extracted 
in biological processes. On the fundamental level, such applications require the development of design principles 
that will allow us to realize powerful and efficient engines that operate between two reservoirs at different 
``temperatures'' (or chemical potentials), and produce useful work with maximum efficiency. In particular, in the 
framework of thermal engines, the question of maximum efficiency has been addressed by the pioneering work 
of Carnot which pointed out that the efficiency of a thermal engine that performs a cycle between two reservoirs 
with temperatures $T_H$ and $T_C$ ($T_H>T_C$) is bounded by the so-called Carnot efficiency $\eta_C=1-
T_C/T_H$ \cite{C24,C85}. Of course, this thermodynamic 
bound is of limited practical importance since the corresponding heat engine must work reversibly, and thus its 
output power is zero. A more practical direction is to identify conditions under which the power of {\it irreversible} 
thermal engines, working under finite-time Carnot cycles, is optimized while their efficiency is still high \cite{CA75,
N57,EKLB10,GMS10,S11,BSC11}. The situation is even more complex when one abandons the convenience of 
macroscopic thermodynamics framework and delves into the challenges of modern nano-devices, where wave 
interferences and thermal fluctuations dominate their performance \cite{GK14,RKSCPCP18,ZBM14,GFMLCD16}. 

\begin{figure}
\includegraphics[width=0.45\textwidth]{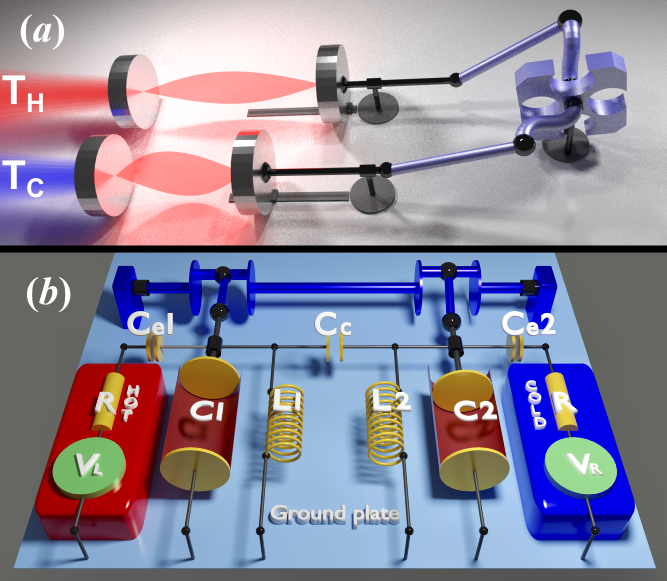}
\caption{Schematic representation of our thermal motor. 
$(a)$ A photonic circuit connected to two thermal baths at different temperatures $T_H>T_C$ is able to divert part 
of the thermal radiative energy into useful work in the form of the motion of a mechanical degree of freedom (MDF) 
described by a rotor. $(b)$ An electromechanical motor consisting of two coupled LC resonators. The capacitance 
plates of the LC circuits are coupled to pistons whose motion is out-of-phase with one another, thus excessing a 
torque to a rotor. When the system operates in the vicinity of EPs and violates specific spatio-temporal symmetries, 
(or the baths are subjected to spectral filtering), the motor operates at optimal performance.
}
\label{fig1}
\end{figure}

A prominent framework where many of these challenges met is in photonics. In this case, the near field thermal 
radiation, emitted from a hot reservoir towards a cold reservoir, can be harvested by an optomechanical circuit 
as a non-conservative ``wind-force". Under its influence, a (slow) mechanical degree of freedom (MDF) undergoes 
a closed path periodic motion. We show that for a long - but finite - driving period of the MDF, these circuits act 
as {\it autonomous radiative motors}, whose extracted power and efficiency are maximized when they are designed 
to operate in a domain of their parameter space which is in the vicinity of an exceptional point (EP) degeneracy. 
The latter signifies a coalescence of the eigenvalues and the corresponding eigenvectors of the effective non-
Hermitian Hamiltonian that describes the coupling of these motors with the thermal reservoirs. We have engineered 
such EPs via a judicious coupling contrast with the reservoirs and we have harvested its influence in the enhancement 
of the performance of the motor, by appropriate manipulation of its spatio-temporal symmetries and of the 
thermal emissivity of the attached reservoirs. Our predictions can guide the design towards optimal operational 
conditions of autonomous motors. Their applicability extends beyond the photonic framework to other platforms 
like electromechanical circuits for harvesting mechanical (e.g. vibrational) ambient noise for power supply of a 
variety of auxiliary systems \cite{BTW06,AS07,MYR08,PI09,GMPN11,WMKW11}.

\section{Mathematical Formulation} 

The system consists of two thermal reservoirs at different temperatures $T_H>T_C$, which are brought 
in contact via a circuit. The latter is coupled to a MDF from which we extract work. For simplicity, we will assume that 
the circuit incorporates two single-mode resonators whose frequencies $\omega_n$ (where $n=1,2)$ are modulated 
by the motion of the MDF. The temperature gradient between the two reservoirs produces a thermal current through 
the circuit that, in turn, exerts a force to the MDF engaging them in slow periodic motion along a given closed trajectory 
${\cal C}$ in a parameter space. The design is chosen in a way that the motion along the path ${\cal C}$ creates 
out-of-phase variations in $\omega_{1,2}$, thus leading to a violation of spatio-temporal symmetries of the structure. 
One possible implementation of the above set-up is in the photonic framework Fig. \ref{fig1}a, while a parallel proposal 
in the electromechanical framework is shown in Fig. \ref{fig1}b. Below, we will mainly use the photonic ``language'' 
associated with Fig. \ref{fig1}a, while we will also have the electromechanical scenario of Fig. \ref{fig1}b in mind.

In typical circumstances, the MDF ${\vec X}=\{X_1,X_2,\cdots,X_M\}$ describes a change in position or angle of the 
mechanical element due to a respective force or torque. For concreteness of our presentation, we will assume that 
$M=2$. The coordinate ${\vec X}$ abides by the Langevin equation
\begin{equation}
{\cal M}{\ddot {\vec X}}(t)  = \Gamma {\dot {\vec X}}(t) +  {\vec F}_{\rm av}  + {\vec \xi}(t) ,
 \label{Langevin}
\end{equation}
where ${\cal M}$ is the generalized inertia tensor, ${\vec \xi}(t)$ is a fluctuating force and $\Gamma$ is the friction 
tensor which satisfies a fluctuation-dissipation relation. In our analysis below, we will assume that the fluctuating force 
${\vec \xi}$ can be neglected due to the large 
inertia of the MDF. Consequently, we can approximate the dynamics of ${\vec X}$ by its mean value ${\vec x}=\left
\langle {\vec X} \right \rangle$, where $\langle\cdot\rangle$ indicates a thermal averaging. Finally, the mean ``force'' 
${\vec F}_{\rm av}$, drives the mechanical rotor diverting energy from the ``photonic'' thermal current to produce 
mechanical work. In the photonic framework (Fig. \ref{fig1}a) ${\vec F}_{\rm av}$ is analogous to the radiation pressure 
associated with the radiation inside the circuit. In the electromechanical framework of Fig. \ref{fig1}b, ${\vec F}_{\rm av}$ 
is associated with a torque acting on the mechanical rotor. 

The interaction between the mechanical part and the radiation is obtained from the variation of the energy inside 
the photonic circuit due to a displacement ${\vec x}$ of the MDF. Specifically, the thermal averaged force is
\begin{equation}
{\vec F}_{\rm av} = -\hbar \left\langle  \Psi^\dagger \frac{\partial H_0 }{\partial {\vec x}} \Psi \right \rangle
= -\hbar \sum_{n,n^\prime} \left(\frac{\partial H_0}{ \partial {\vec x}} \right)_{n n^\prime} \left\langle \psi_n(t)^* 
\psi_{n^\prime}(t)\right \rangle.
\label{Fav}
\end{equation}
where $\Psi =\left(\psi_1,\cdots,\psi_N\right)^T$, $\psi_n$ is the field amplitude at the $n-$th resonator of the 
circuit ($\hbar \omega_n |\psi_n|^2$ represents the energy density in the $n-$th mode/resonator), and $H_0(
{\vec x})$ is the effective Hamiltonian of the circuit that provides a description of the dynamics of the radiation 
field in the single-mode resonators. The dynamics of the open system (circuit coupled with reservoirs) is described 
in terms of a temporal coupled-mode theory (CMT) \cite{H00}
\begin{eqnarray}
\im \frac{d \Psi (t) }{dt}  &=& H_{\rm eff} \Psi(t) + \im D^T \theta^{(+)}(t) ;  
H_{\rm eff}=H_0({\vec x}) -\im \frac{D^T D}{2}   \nonumber\\
\theta^{(-)} &=& -\theta^{(+)} + D \Psi ; 
\label{CMT}
\end{eqnarray} 
where the matrix $D$, with elements $D_{n,\alpha} = \sqrt{2\gamma_{\alpha}} \delta_{n,\alpha}$, describes the 
coupling of the circuit with the thermal baths. The thermal excitations from (towards) the $\alpha$-th reservoir are 
given by the incoming (outgoing) complex fields $\theta^{(+)}$ [$\theta^{(-)}$]. In the frequency domain (using 
the Fourier transform $f(t)=\int_{0}^{\infty} f(\omega) e^{-\im \omega t}d \omega$), the amplitudes $\theta^{(+)
}_\alpha(\omega)$ satisfy the relation
\begin{equation}
\label{fluct}
\left\langle  [\theta^{(+)}_{\alpha^\prime}(\omega^\prime)]^* \theta^{(+)}_{\alpha}(\omega) \right \rangle ={1\over 
2\pi}{\tilde \Theta}_{\alpha}(\omega)\delta(\omega-\omega')\delta_{\alpha,\alpha^\prime}
\end{equation} 
where ${\tilde \Theta}(\omega)=\Phi(\omega)\cdot \Theta(\omega)$, with $\Phi(\omega)$ being a noise filter function 
and $\Theta_{\alpha}(\omega)=\left\{ \exp\left[\hbar\omega/(k_B T_\alpha)\right]-1 \right\}^{-1}$ is the Bose-Einstein 
statistics describing the mean number of photons which are emitted from reservoir $\alpha$ with frequency $\omega$. 
Finally $T_{\alpha}$ is the temperature of the $\alpha$-th reservoir.

\section{Work}

We assume that the dynamics of the MDF Eq. (\ref{Langevin}) occurs on time-scales much larger 
than the ones associated with the field dynamics Eq. (\ref{CMT}). Under this assumption, we can invoke the Born
-Oppenheimer approximation and obtain the work performed by the motor along the path $C$ as\cite{BC19,BRO13,
DMT09,BKVEO11,FPB17,FBP15,FPB19} (see supplement)
\begin{eqnarray}
 W &=&  \int_0^\infty \frac{d\omega}{2\pi} \sum_\alpha{\tilde \Theta}_\alpha(\omega) P_\alpha(\omega); \label{work}\\
 P_\alpha&=& \frac{\hbar}{\im}
 \oint_C d {\vec x} \left[ (S^{\vec x})^\dagger \nabla_{\vec x} S^{\vec x} \right]_{\alpha,\alpha}
 =\hbar \oint _C R^{\vec x} \nabla_{\vec x}\alpha^{\vec x}d {\vec x}. \nonumber 
\end{eqnarray}
where $S^{\vec x} (\omega)= -I_{N_\alpha} + \im D G^{\vec x}(\omega) D^T$ is the unitary instantaneous 
scattering matrix, and $G^{\vec x}=[\omega I_N - H_{\rm eff}(x)]^{-1}$ is the Green's function associated with 
the effective Hamiltonian $H_{\rm eff}$ ($I_m$ is the $m\times m$ identity matrix). In Eq. (\ref{work}), the kernel 
$P_\alpha(\omega)$ indicates the spectral response of the system at a frequency $\omega$. Since $P_\alpha(
\omega)$ only involves a parametric integral along the path $C$,  it is a geometric quantity\cite{BS20,BTTOFA20}. 
It turns out that for the two-reservoir setup of Fig. \ref{fig1}, $P_\alpha$ can be written only in terms of the 
reflectance $R^{\vec x}$ and the corresponding reflection phase $\alpha^{\vec x}$ (see the right part of Eq. (
\ref{work})). As a matter of convention, a positive $W$ in Eq. (\ref{work}) indicates that the dynamics of ${\vec x}$ 
follows the positive direction of the path ${\cal C}$. 

An analytically useful expression of $P_{\alpha}$ is achieved by substituting in Eq. (\ref{work}) the scattering matrix 
in terms of the Green's function. We get
\begin{eqnarray}
\label{P_Green}
P_\alpha=   \iint_A \small{\frac{\partial \omega_p}{\partial x_p} \frac{\partial\omega_q}{\partial x_q} } 
    {\cal W}_{\alpha} d x_p d x_q,
\end{eqnarray}
where we have used that $\left(\frac{\partial H_0}{\partial x_p}\right)_{n,m} = \frac{\partial \omega_n}{\partial x_p}
\delta_{n,m} \delta_{n,p}$, and we have introduced the work density per unit area as
\begin{equation}
\label{W_Green}
 {\cal W}_\alpha =\lim_{A\rightarrow 0} P_\alpha/A = 4 \gamma_\alpha 
\hbar{\cal R}e \left( G_{p\alpha}^* G_{pq} G_{q\alpha} - G_{q\alpha}^* G_{qp} G_{p\alpha} \right),
\end{equation}
with $A= \iint_A \frac{\partial \omega_p}{\partial x_p} \frac{\partial \omega_q}{\partial x_q}d x_p d x_q\rightarrow0$. 

Direct inspection of Eq. (\ref{work}) allow us to establish the following two conditions for the implementation of 
our proposal as a motor: (a) the force has to be non-conservative which means that the $\nabla_{\vec x} \times 
\left[ (S^{\vec x})^\dagger \nabla_{\vec x} S^{\vec x} \right]_{\alpha,\alpha}\ne 0$, and (b) the closed path ${\cal 
C}$ must enclose a non-zero area in the parameter space $\{x_1,x_2\}$. A bi-product of the last condition is 
that variations of $x_1,x_2$ with a phase difference $0$ or $\pi$ cannot produce work.

\begin{figure*}
\includegraphics[width=0.9\textwidth]{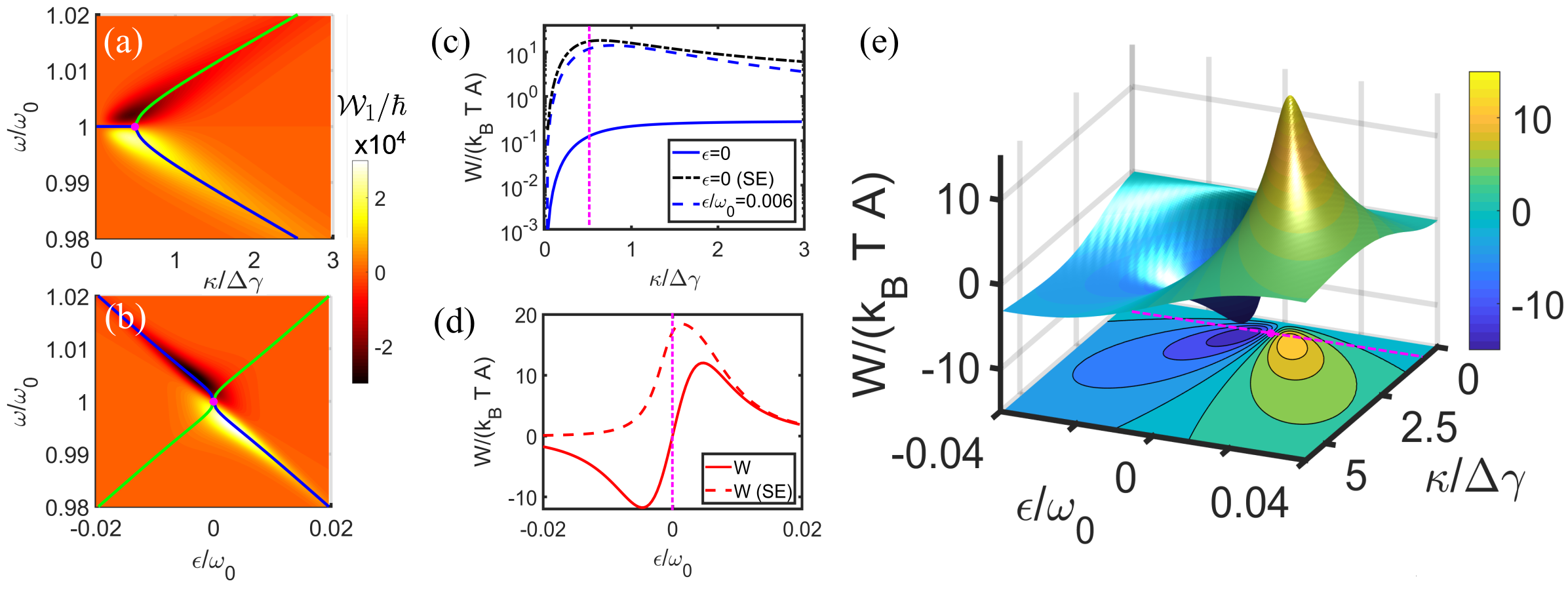}
\caption{{\bf Work in the proximity of an EP.} $(a)$ and $(b)$ The work density (color scale) as a function of the 
coupling constant $\kappa$ and frequency $\omega$ of the incident radiation. Blue and green solid lines are 
the eigenfrequencies of $H_{\rm eff}$ which they demonstrate an EP degeneracy (indicated by the magenta point) 
at $\kappa / \Delta \gamma = 0.5$ in $(a)$ and at $\epsilon=0$ in $(b)$. $(a)$ In the absence of a resonance 
detuning, i.e., $\epsilon=0$, the work density is maximized (minimized) close to the EP for frequencies $\omega
<{\rm Re}(\omega_{\rm EP})$ [$\omega>{\rm Re}(\omega_{\rm EP})$]. $(b)$ A nonzero detuning, i.e., $\epsilon
\approx0.006\omega_0$, breaks the symmetry along the frequency axis. Here $\kappa/\Delta\gamma=0.5$. $(c)$ 
The work $W$ for $\epsilon=0$ ($\epsilon=0.006\omega_0$) is indicated with a solid (dashed) blue line respectively. 
At the same subfigure, we report (short dashed black line) the work in the case where we have introduced a spectral 
filtering function $\Phi(\omega)=H(\omega-\omega_0)$ ($H(x)$ is the Heaviside function). $(d)$ Work with spectrally 
engineered reservoirs for $\epsilon=0$ (solid red line) and $\epsilon\approx0.006\omega_0$ (dashed red line). 
From subfigures $(c)$ and $(d)$ we conclude that a perturbation that explicitly violates the (pseudo)-${\cal PT}$
-symmetry (e.g. due to resonant detuning) or/and appropriate spectral filtering of the bath maximizes the work 
$W$ in the parameter-space domain which is in the proximity to the EP. In the above examples, the resulting 
enhancement factor is $\sim200$-fold as opposed to the unperturbed/unfiltered case. {\bf Total work} $(e)$ We 
report the work $W$ as a function of the coupling constant $\kappa$ and the perturbation $\epsilon$. The EP 
parameters are indicated with a magenta dot in the colormap (the magenta line indicates the value of 
$\kappa_{EP}$). In the vicinity of the EP, the work becomes extreme (positive or negative). The CMT parameters 
used in these calculations are: $\omega_0=200\times 10^{12} rad/s$, $T_1=10^9K$, $T_2=3K$,  and the rest 
of parameters are in units of $\omega_0$:  $\gamma_1=0.01$, $\gamma_2=0.002$, $\bar \omega_1=\bar 
\omega_2 =1$.
}
\label{fig2}
\end{figure*}

\section{Engineering EP degeneracies} 

In the frequency range near an EP-degeneracy the resolvent of the effective 
Hamiltonian $H_{\rm eff}$ can be approximated by a $2\times2$ subspace involving only the resonant modes 
associated with the EP. We therefore consider a minimal model consisting of two coupled modes with resonant 
frequencies $\omega_1,\omega_2$. Alternatively, one can consider, as a concrete example, the set-up of Fig. 
\ref{fig1}a. The system consists of two single-mode resonators coupled asymmetrically to two reservoirs at 
temperatures $T_{\alpha=1}=T_H$ and $T_{\alpha=2}=T_C$. The effective Hamiltonian of such a reduced system 
reads
\begin{eqnarray}
  H_{\rm eff}=
  \begin{pmatrix}
    \omega_1 -\im \gamma_1  &  \kappa    \\
    \kappa     &  \omega_2 -\im \gamma_2  
  \end{pmatrix}
\label{Heff}
\end{eqnarray}
where $\kappa$ describes the coupling between the two modes and $\gamma_1,\gamma_2$ are the (asymmetric) 
decay rates of the two modes due to their coupling with the two reservoirs. The spectrum of $H_{\rm eff}$ is 
$\omega_{\pm} = \omega_0 -\im \gamma_0 \pm \frac{1}{2}\sqrt{ \left( \Delta \omega -\im \Delta \gamma \right)^2 +
4\kappa^2 }$, where $\omega_0=\frac{\omega_1 + \omega_2}{2}$, $\gamma_0=\frac{\gamma_1+\gamma_2}{2}$ 
and $\Delta \omega = \omega_1-\omega_2$ and $\Delta \gamma=\gamma_1-\gamma_2\ne0$. The corresponding 
(non-normalized) eigenvectors are $u_{1,2}= \left( 2\kappa, -\Delta\omega + \im\Delta\gamma \pm \sqrt{ \left( \Delta 
\omega -\im \Delta \gamma \right)^2 +4\kappa^2 }\right)$. It is easy to show that when $\Delta \omega=\Delta 
\omega_{\rm EP}=0$ and $\kappa_{\rm EP}=\Delta\gamma/2$ the system supports an EP degeneracy with 
$\omega_+ = \omega_-= \omega_{\rm EP}=\omega_0 -\im \gamma_0$. 

In fact, under the condition $\Delta \omega =\Delta \omega_{\rm 
EP}$, the Hamiltonian Eq. (\ref{Heff}) respects a (pseudo-)parity-time (${\cal PT}$) symmetry that reveals itself 
after renormalizing the losses with respect to their mean value $\gamma_0$ \cite{GSDMVASC10}. Below we will 
be discussing in detail two distinct scenarios involving perturbations around the EP that violate this (pseudo-)
${\cal PT}$-symmetry either spontaneously or explicitly. We will show that each of these cases affects in a dramatically 
different manner the characteristic features of the work density ${\cal W}_{\alpha}$.

\section{Work density in the presence of an EP} 

We analyze the extracted work density of the motor when the center 
of the modulation cycle is in the proximity of an EP. To this end, we consider a modulation cycle 
${\cal C}$ associated with changes of the resonant frequencies being $\omega_{n} = \omega_{0} \left(1+ \delta \cos(x_n+\phi_{n})\right)]-(-1)^n\epsilon$, where $\epsilon$ describes a resonance detuning that displace 
the unmodulated system Eq. (\ref{Heff}) from the EP by violating explicitly its (pseudo-) ${\cal PT}$-symmetry. In 
order to satisfy the criteria for non-zero work, we have assumed that the two resonances are modulated out of 
phase i.e. $\phi_{1}=\pi/2, \phi_2=0$. For such a modulation scenario, the associated enclosed area in the parameter 
space $\left(\omega_{1}(x_1),\omega_{2}(x_2)\right)/\omega_0$ is $A=\pi \delta^2$. 

Next, we assume a generic perturbation $p$ which displaces the center of the modulation cycle with respect to the 
EP. Using Eq. (\ref{W_Green}), we have evaluated the work density ${\cal W}_{\alpha}$ in terms of the Green's 
function $G^{\vec x}$. In fact, for the $2\times2$ case, the calculations for the Green's function can be carried out 
explicitly for any perturbation, giving 
\begin{eqnarray}
  G^{\vec x}&=& \frac{1}{D} 
  \begin{pmatrix}
    \omega-(\omega_2 -\im \gamma_2)  &  -\kappa    \\
    -\kappa     &  \omega-(\omega_1 -\im \gamma_1)  
  \end{pmatrix}\nonumber\\
&\approx& \frac{A}{\omega-\omega_{EP}}  + \frac{B}{(\omega-\omega_{EP})^2},
  \label{GreenF}
\end{eqnarray}
where $D=[ \omega-(\omega_2 -\im \gamma_2)][\omega-(\omega_1 -\im \gamma_1)] -\kappa^2$. In the above 
expression, the generic perturbation $p$ is ``hidden'' in the parameters that define $H_{\rm eff}$ e.g. in the 
frequencies $\omega_{1,2}=\omega_{1,2}(p)$ and/or the coupling $\kappa=\kappa(p)$ between the two resonant 
modes. When $p\rightarrow 0$ the Green's function can be approximated with the last expression, where $A$ 
and $B$ are frequency-independent matrices (see methods). It turns out that the functional dependence of ${\cal 
W}_{\alpha}$ on $\omega$, in the vicinity of the EP, is dramatically affected by the presence of the square-Lorentzian 
term on the last part of Eq. (\ref{GreenF}). This unique spectral feature is a consequence of the degeneracy of 
the eigenvectors of $H_{\rm eff}$ at the EP. Furthermore, a squared Lorentzian lineshape implies a narrower 
emission/absorption peak and greater resonant enhancement in comparison with a non-degenerate resonance 
at the same complex frequency. We will show that the competition between the two terms appearing at the right 
equality of Eq. (\ref{GreenF}) determines the conditions under which ${\cal W}_{\alpha}$ acquires its maximum 
value (see below).

A more elaborated treatment can extend the above analysis of $G^{\vec x}$, in order to include any number of 
modes, by using a degenerate perturbation theory that takes into consideration the singular nature of EPs. In this 
case, the standard modal decomposition of the Green's function is not applicable since the bi-orthogonal 
eigenvectors of $H_{\rm eff}$ do not span the Hilbert space. Instead, one has to complete the eigenvectors of 
$H_{\rm eff}$ into a basis by introducing the associated Jordan vectors \cite{PZMHHRSJ17}. Following this approach, 
we can recover the last expression of $G^{\vec x}$ in Eq. (\ref{GreenF}). 

Substituting the expression for the Green's function back in Eqs. (\ref{P_Green},\ref{W_Green}) we get that:
\begin{eqnarray}
  {\cal W}_1 &=& 4\hbar\gamma_1 \gamma_2 {\rm Re}\left[ \frac{-2\im \kappa^2 D^*}{|D|^4} \right]  \label{W_general}\\
   &=& \frac{-8\hbar\gamma_1\gamma_2\kappa^2\left(2\gamma_0(\omega-\omega_0) +\epsilon\Delta\gamma \right)}
   {\left\{ \left[(\omega-\omega_0)^2 - \gamma_0^2-\epsilon^2 +c  \right]^2 + \left[2\gamma_0(\omega-\omega_0) + \epsilon\Delta\gamma\right]^2 \right\}^2} \nonumber
\end{eqnarray}
where for the evaluation of the contour integral in Eq. (\ref{P_Green}) we have explicitly written $\omega_{1,2}$ 
in terms of the parameters $\omega_0$ and $\epsilon$ that define the position of the path ${\cal C}$. The 
constant $c=(\Delta\gamma/2)^2-\kappa^2$ and/or with the detuning $\epsilon$ indicate the degree of deviation 
from the EP.

Let us exploit further Eq. (\ref{W_general}) by considering two specific examples corresponding to perturbations 
that preserve/violate the pseudo-${\cal PT}$ symmetry of the effective unmodulated Hamiltonian $H_{\rm eff}$. In 
the first case, we displace the system away from the EP by varying the coupling $\kappa\ne\kappa_{\rm EP}$ while 
keeping $\epsilon=0$. We find that the work density takes the form
\begin{equation}
{\cal W}_1=-\frac{16\hbar\gamma_1 \gamma_2 \kappa^2 \gamma_0 (\omega-\omega_0)}{ \left\{\left[(\omega-
\omega_0)^2 + \gamma_0^2 + c\right]^2 -4\gamma_0^2 c  \right\}^2}.
\label{EP_W}
\end{equation}
In fact, by considering 
the EP condition $c=0$ we are able to identify in the denominator of ${\cal W}_1$ above, the signature of the 
square-Lorentzian anomaly associated with the collapse of the eigenvector basis. Equation (\ref{EP_W}) allows 
us to conclude that ${\cal W}_{1}$ is non-monotonic and antisymmetric with respect to the EP resonance 
frequency axis $\omega=\omega_0$ for all $\kappa$-values. Furthermore, ${\cal W}_1(\omega=\omega_0)
=0={\cal W}_1(\omega\rightarrow \pm \infty)$ while its extrema occur in the vicinity of the EP (see the 
filled magenta circle) at $\omega=\omega_0\pm \sqrt{1\over 7} \gamma_0$, see Fig. \ref{fig2}a.

The situation is dramatically different when we choose to perturb the system away from the EP using a parameter 
that enforces an explicit (pseudo-)${\cal PT}$-symmetry violation of the unmodulated Hamiltonian $H_{\rm eff}$. 
An example case is when the resonances of the two coupled modes are detuned by $\epsilon$. In this case, the 
diagonal elements of $H_{\rm eff}$ take the form $\omega_{1,2} = \bar \omega_{1,2} \pm \epsilon +  \omega_0
\delta \cos(x+\phi_{1,2})$. Furthermore, the work density does not have a definite symmetry with respect to $(\omega
-\omega_0)$. To be concrete, we consider the particular case $\kappa=\kappa_{EP}=\Delta\gamma/2$ for which the 
work density is
\begin{equation}
  {\cal W}_1=-\frac{8\hbar\gamma_1 \gamma_2}{|D|^4} \left(\frac{\Delta\gamma}{2}\right)^2  
  \left[ 2 \gamma_0 (\omega-\omega_0) + \epsilon \Delta\gamma\right]  
  \label{W_epsi}
\end{equation}
where now the denominator takes the form $|D|^4=\left\{  \left[ (\omega-\omega_0)^2 - \gamma_0^2 -\epsilon^2 
\right]^2 + \left[ 2\gamma_0(\omega-\omega_0) +\epsilon\Delta\gamma \right]^2 \right\}^2$ demonstrating the 
traces of the square-Lorentzian anomaly. The latter is better appreciated 
in the limit of $\epsilon=0$ (EP condition). For $\epsilon\ll\omega_0$, we can further expand up to leading order 
in $\epsilon$ the denominator and get 
\begin{eqnarray}
  {\cal W}_1&\approx&-\frac{8\hbar\gamma_1 \gamma_2}{|\omega-\omega_{EP}|^8} \left(\frac{\Delta\gamma}{2}\right)^2  
\left[2 \gamma_0(\omega-\omega_0)+  \right. \label{W_pert}\notag\\
  & + & \left.  \epsilon\Delta\gamma\left(1 - \frac{16(\omega-\omega_0)^2\gamma_0^2}
{|\omega-\omega_{EP}|^{4}} \right)  \right],  
\end{eqnarray}
where the term associated with the perturbation $\epsilon$ is an even function in $(\omega-\omega_0)$. We 
conclude, therefore, that the work density ${\cal W}_{\alpha}$ loses the parity as soon as $\epsilon$ is turned 
on, see also Fig. \ref{fig2}b. Below we will be discussing the consequences of such effect in the power extraction of 
the autonomous motor.

\section{Work in the presence of EP}

We are now ready to exploit the properties of ${\cal W}_{\alpha}$ for the design 
of autonomous motors with optimal performance. To this end, we remind that the extracted work $W$ is essentially 
the frequency integral of ${\cal W}_{\alpha}$, weighted with the function ${\tilde \Theta}_\alpha(\omega)$, see 
Eq. (\ref{work}).
 
Let us first discuss the family of perturbations that preserve the (pseudo-)${\cal PT}$-symmetry of the unmodulated 
effective Hamiltonian. In this case, the antisymmetric form of the work density ${\cal W}_\alpha$ with respect to the 
$\omega_{0}-$axis, results in a near-zero total work, see Fig. \ref{fig2}c. The slight deviation from zero (towards positive 
$W>0$) is due to the fact that Eq. (\ref{work}) involves a product of ${\cal W}_\alpha$ with $\tilde{\Theta} (\omega)$ 
which slightly de-symmetrizes the integrand towards smaller frequencies (see continuous blue line). We can revert 
the situation by introducing a spectral filtering function $\Phi(\omega)$ which enhances the unbalance contribution 
of positive and negative work densities in the integral of Eq. (\ref{work}). The resulting extracted work, for the example 
case of a filter function $\Phi(\omega)=H(\omega-\omega_0)$, is reported in Fig. \ref{fig2}c with a black dashed line 
($H(x)$ is the Heaviside function). Our results indicate that such a spectral filtering approach can lead to an increase 
in $W$ which is higher by two orders of magnitude with respect to the unfiltered case. The same data indicate that 
the maximum work occurs in the vicinity of the EP where ${\cal W}_{\alpha}$ acquires its maximum value (violet vertical 
line) and where the de-symmetrization strategy via spectral filtering is more impactful.  

An alternative way to induce an asymmetric integrand in Eq. (\ref{work}) is by perturbing the system away 
from the EP via a perturbation that will explicitly violate the (pseudo)-${\cal PT}$ symmetry of the unmodulated
effective Hamiltonian. In the previous section, we have identified one such perturbation being the frequency 
detuning $\epsilon$ between the two resonators. In this case, the work density itself becomes asymmetric 
(see Fig. \ref{fig2}b), leading to a frequency integral Eq. (\ref{work}) which is different from zero. In fact, the 
maximum $W$ occurring in the proximity of $\kappa_{\rm EP}$, is again two orders of magnitude enhanced 
in comparison to the $\epsilon=0$-case, see the blue dashed line in Fig. \ref{fig2}c.

The enhancement of the extracted work $W$ via engineered perturbations that violate the (pseudo)-${\cal PT}$-
symmetry of the motor is better appreciated in Fig. \ref{fig2}d. Here, we report the extracted work $W$ (for fixed 
$\kappa=\kappa_{\rm EP}$) for both spectrally unfiltered/filtered noise versus the perturbation $\epsilon$. For 
the unfiltered case (solid red line), we find that in the vicinity of the EP the total work is proportional to $\epsilon$, 
a relation that it is a direct consequence of the expansion Eq. (\ref{W_pert}) for the work density. Specifically, 
assuming for simplicity that $\Theta_1 (\omega)\approx\Theta_1(\omega_0)$, the integration over $\omega$ leads 
to the conclusion that $W\approx A \Theta_1 (\omega_0)\int d\omega/(2\pi) {\cal W}_1\propto \epsilon$. The same 
argument applies also in the case of spectral filtering with $\Phi(\omega)=H(\omega-\omega_0)$ (see dashed red 
line). In both cases, the extreme work $W_{\rm max}$ occurs at perturbation strengths $\epsilon_{\rm max}$ 
in the vicinity of the EP, where the linear approximation Eq. (\ref{W_pert}) breaks down. An additional conclusion 
that we extract from the above analysis is that the spectral filtering method (combined with perturbations that 
violate the (pseudo)-${\cal PT}$-symmetry lead to a slightly (two-fold) increase of the extracted work as compared
to the unfiltered case (see solid red line). 

A panorama of the extracted work $W$ versus $\epsilon$ and $\kappa$ is shown in Fig. \ref{fig2}e. Here we
report only the unfiltered case i.e. $\Phi(\omega)=1$. The data demonstrate nicely that the extreme value of 
the extracted work occurs in the vicinity of $(\epsilon, \kappa)=(0,\kappa_{\rm EP})$ where the EP is located. 
The case of spectral filtering with a function $\Phi(\omega)$ (e.g. $\Phi(\omega)=H(\omega-\omega_0)$) shows 
the same qualitative features (with the only difference that $W$ is flat in the negative $\epsilon$ semi-plane 
due to the specific filter function) and therefore is not reported here.

\begin{figure}
 \includegraphics[width=0.5\textwidth]{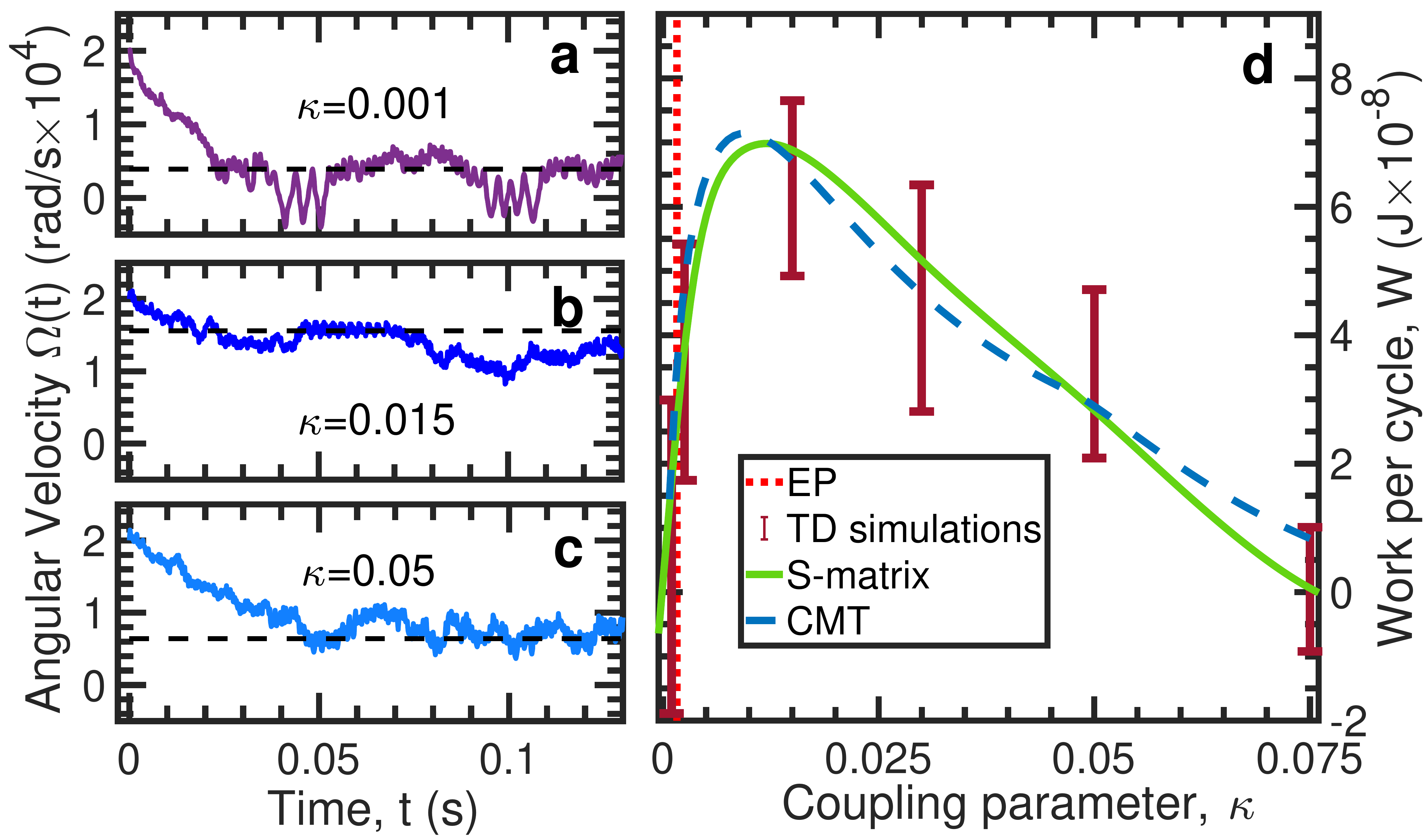}
\caption{ (a-c) The dynamics of the angular velocity $\Omega(t)$ for some representative values of the
coupling coefficient $\kappa$ whose terminal velocity determines the work delivered by the photonic circuit. 
(d) Work performed by the two-resonator circuit setup versus the coupling parameter $\kappa$. The numerical 
evaluation for the work (dots with error bars) is based on the value of the terminal angular velocity, see Eq. 
(\ref{W_EMS}). The TD simulations match nicely the theoretical predictions for the work (green line) given by Eq. 
\ref{work}. The blue dashed line reports the work predicted by the CMT modeling, see Eqs. (\ref{CMT},\ref{Heff}).\cite{CMTpar} 
The vertical red dotted line indicates the position of the EP.
}
\label{fig3}
\end{figure}

\section{Time Domain Simulations and Implementation using Electromechanical Systems} 

We validate the above 
proposal by performing time-domain (TD) simulations using COMSOL software\cite{COMSOL} with a realistic 
electromechanical system, see Fig. \ref{fig1}b. The setup consists of a pair of capacitively coupled resonators 
with impedance $Z_0=70$ Ohm tuned at different frequencies, $\omega_{1,2}=\omega_0\pm \epsilon$, 
which enforces violation of the (pseudo-)${\cal PT}$-symmetry of the unmodulated system. In our simulations, 
we have considered that $\omega_0 = 2\pi f_0$, with $f_0=1$ MHz, and $\epsilon = 0.0488\cdot\omega_0$. 
The capacitors $C_{1,2}$ are considered as a pair of conductive plates separated by a median air gap $d_0=
\frac{e_0\cdot A }{C_0}$, where $e$ is the vacuum permittivity and $C_0 = \frac{1}{Z_0\cdot \omega_0}$ is a 
median capacitance. The upper plates of the capacitors are assumed to be attached to a wheel (the MDF) of 
radius $r=d_0/10$ in a way that during the wheel rotation with angular velocity $\Omega$ the plates will undergo 
a motion described by the displacements $d_{1,2} = d_0+r\cdot \cos(\phi_{1,2})$ with $\phi_1= \Omega t$ and 
$\phi_2=\phi_1+\pi/2$. The wheel is assumed to have mass $m=1$ g, moment of inertia $I=0.5\cdot m r^2 = 
7.58\cdot 10^{-15}$ kg$\cdot$ m$^2$, and experiences friction with the ambient medium with friction coefficient 
$\Gamma = 2.5\cdot10^{-13}$ N$\cdot$m $\cdot$s/rad. The coupling capacitance between the two $LC$ 
resonators is $C_c=2\kappa\cdot C_0$, where the coupling coefficient $\kappa$ is a tunable parameter of the 
simulations. The left/right resonators are coupled to a hot/cold baths via capacitors $C_{e1}=0.1\cdot C_0$, 
and $C_{e2}=0.03\cdot C_0$ respectively, which yields the following value for the critical coupling $\kappa_{EP} 
= 0.001625$ (red dotted line on Fig. \ref{fig3}d). 

To enhance further the extracted work from the MDF  we have introduced, in addition to the detuning $\epsilon$, 
spectral filtering of the thermal baths. Specifically, the hot bath is producing a noise signal consisting of $200$ 
spectrally uniformly distributed harmonics $V(t) = V_0\cdot \sum_{i=1}^{200}  \sin \left(\tilde\omega_i\cdot  t+
\varphi_i\right)$, where $V_0 = 1$V is the amplitude of the noise, $\tilde\omega_i$ is a frequency of each noise 
harmonic, and  $\varphi_i$ - is a random phase shift. The lower frequency of the noise considered in the simulations 
is $\tilde\omega_1= 2\pi\cdot 0.85$ MHz with an upper limit of $\tilde\omega_{200}= 2\pi\cdot 1.1$ MHz. 

In the simulations the wheel is given an initial angular velocity $\Omega_0=2.5\cdot10^4$ rad/s. Its angular 
velocity is monitored as a function of time until it saturates at a certain value $\Omega_s$. From here, we evaluate 
the work per cycle via the relation 
\begin{equation}
\label{W_EMS}
W_{TD} = \int_{0}^{2\pi} \tau(x) dx=2\pi\cdot \Gamma \cdot \Omega_s,
\end{equation}
where $\tau$ is the torque produced by the capacitor plates on the wheel and $x$ is the angular displacement. The 
subindex TD indicates that the evaluated work is extracted from our time-domain simulations. 

In Figs. \ref{fig3}a-c we show the transient dynamics of the angular velocity $\Omega(t)$ for three typical coupling 
constants $\kappa$. Notice that in some cases (e.g Fig. \ref{fig3}a) the angular velocity $\Omega(t)$ acquires 
negative values indicating that the wheel rotates opposite to the direction of the closed path ${\cal C}$. We find 
that in the long time limit the MDF reaches a terminal angular velocity 
$\Omega(t\rightarrow \infty)\equiv \Omega_s^{TD}$ which can be used in Eq. (\ref{W_EMS}) for the numerical evaluation 
of $W_{TD}$. In each of the subfigures \ref{fig3}a-c, we are also indicating (see dashed black line), the theoretical values 
of the saturation velocity $\Omega_s$. The latter has been extracted via Eq. (\ref{W_EMS}), where the work $W$ on the 
left-hand-side has been calculated using Eq. (\ref{work}). For the theoretical evaluation of $W$, we have extracted the 
elements of the instantaneous $S$-matrix of the circuit using a frequency domain analysis of COMSOL\cite{COMSOL}. 

In Fig. \ref{fig3}d we report a summary of the extracted $W_{TD}$ versus the coupling constants $\kappa$. The error 
bars reflect the fluctuations in the numerical evaluation of $\Omega_s^{TD}$ and are extracted from the temporal analysis 
of $\Omega(t)$ as $\Omega_{min/max}^{TD} = min/max\left(\Omega(t\in[t_1,t_{max}])\right)$, where $t_1$ is the time during 
which $\Omega(t)$ reaches the theoretical value of $\Omega_s$ for the first time for a given value of $\kappa$; $t_{max}=0.13 s$ 
- is  a maximum time used in a TD analysis. At the same figure, we are also plotting the theoretical predictions for the 
work $W$ (green line) that have been derived using Eq. (\ref{work}) with instantaneous scattering matrix elements given 
by the COMSOL frequency analysis of the electromechanical system. Finally, at the same figure, we are presenting the 
predictions of the CMT modeling of Eqs. (\ref{CMT},\ref{Heff}). In the latter case, the various parameters (coupling, 
resonance frequencies, linewidths, etc.) of the CMT model have been extracted from the transmission spectrum of the 
electronic circuit (see methods). The nice agreement between CMT and TD simulations confirm the validity of our CMT 
modeling and establishes the influence of the EP protocols in extracting maximum work from thermal autonomous motors.

\section{Efficiency} 

The temperature gradient between the two thermal reservoirs induces a thermal current that 
goes through the motor. Part of the associated input power is dissipated due to friction, resulting in a reduction 
in the amount of {\it usable output power}\cite{FBP15}. The latter can be used e.g. for lifting a weight or charging 
a capacitor. The usable output power is 
\begin{equation}
P_{\rm out}=\frac{\Omega_s}{2\pi} W - \frac{\Omega_s}{2\pi}\Gamma \int_0^{ \frac{2\pi}{\Omega_s} } {\dot x}^2 dt 
\approx \frac{\Omega_s}{2\pi} W - \Gamma \Omega_s^2,
\label{P_u}
\end{equation}
where we have assumed that the MDF has large inertia, forcing the rotor to move with terminal velocity ${\dot 
x} \approx \Omega_s$. The optimal terminal angular velocity that maximizes the usable work is dictated by the 
parameters of the setup, and can be found from Eq. (\ref{P_u}) to be $\Omega_s^*=W/(4\pi \Gamma)$ leading 
to $P_{\rm out}^{*}=\left(\frac{W}{4\pi}\right)^2 {1\over\Gamma}$, which is half of the total ``frictionless'' power 
$\left(\frac{W}{4\pi}\right)^2{2\over\Gamma}$. For circuit parameters such that $\Omega_s\gg\Omega_s^*$, the 
motor dissipates most of the incident energy while in the other limiting case where $\Omega_s\ll \Omega_s^*$ 
the friction can be neglected but the device does not generate much power. In both limits, the usable output power 
is nearly zero. 

\begin{figure}
\includegraphics[width=0.45\textwidth]{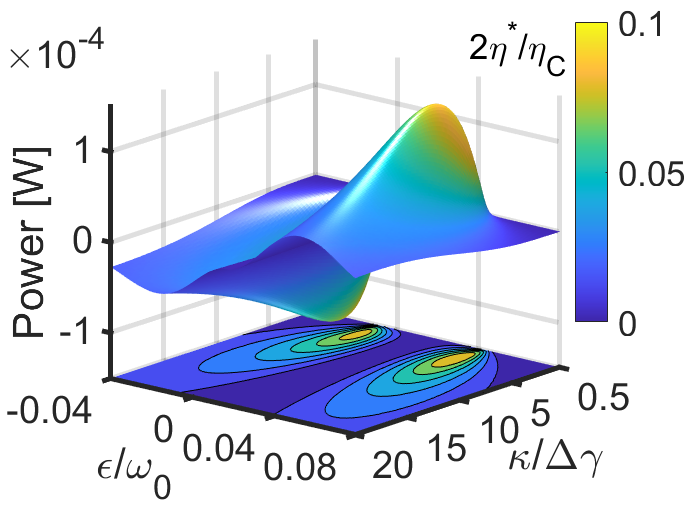}
\caption{Maximum power $P_{\rm out}^*$ (z-axis) and efficiency $\eta^*$ normalized with respect to the maximum 
efficiency $\eta_C/2$ (color-scale) at optimal operational conditions corresponding to $\Omega_s=\Omega_s^*$. 
These quantities are plotted as a function of the perturbation parameters $\kappa$ and $\epsilon$ for a fixed 
temperature gradient. The former perturbation
respects the pseudo-${\cal PT}$-symmetric nature of the unmodulated system while the latter violates this symmetry. 
In these extensive simulations, we have used the CMT modeling with parameters associated with the circuit setup 
(see the previous section).}
\label{fig4}
\end{figure}

It is, therefore, useful to quantify the performance of an autonomous motor by introducing its efficiency $\eta$. 
The latter is defined as the ratio of the net usable average output power $P_{\rm out}$ that is extracted from the 
motor during one period of the cycle $2\pi/\Omega_s^*$ when it operates under optimal conditions (i.e. $\Omega_s
=\Omega_s^*$), to the total input power $P_{\rm in}$ delivered to the photonic circuit. Specifically:
\begin{equation}
\label{eta}
\eta^* \equiv \frac{P_{\rm out}^{*}}{P_{\rm in}}, \quad P_{\rm in} \approx \bar{I}_{b} + \bar{I}_{p}
\end{equation}
where for the evaluation of $P_{\rm in}$ we have also considered the fact that the slow variation of the photonic 
network's parameters induces a pumping energy current $\bar{I}_p$ in addition to the energy current $\bar{I}_b$ 
due to the temperature bias\cite{LAESK19}. Both currents above are measured at the hot reservoirs. Since typically 
$\bar{I}_p\ll \bar{I}_b$ we can omit the pumped current from the denominator while we can substitute in Eq. (\ref{eta}) 
the maximum usable power as $P_{\rm out}^*\sim W^2/\Gamma$. Therefore $\eta^* \sim ({W\over\bar{I}_b}){W\over
\Gamma}$, which suggests that the maximum $\eta^*\le\frac{\eta_C}{2}$ might be expected in the parameter domain 
where $W$ acquires its maximum values, see Figs. \ref{fig2},\ref{fig3}.

An efficient way to test the above expectations of the performance of our EP-influenced motor is by simultaneous 
evaluation of its efficiency Eq. (\ref{eta}) together with the corresponding power $P_{\rm out}^{*}$. These quantities 
are plotted in Fig. \ref{fig4} as a function of the perturbation parameters $\kappa, \epsilon$ associated with the 
coupling and the resonance detuning between the two LC resonators of the electromechanical system of the previous 
section. For these calculations, we have used the CMT modeling with parameters that reproduce the results of the 
direct TD simulations of COMSOL for the electromechanical motor\cite{CMTpar} (see Fig. \ref{fig3}d). Furthermore, 
we have ensured that the angular frequency $\Omega_s^*$ is small enough such that the Born-Oppenheimer 
approximation is valid. From Fig. \ref{fig4} we see that both $\eta^*$ and $P_{\rm out}^*$ acquire their maximum 
values at the vicinity of the EP --albeit at slightly different $(\kappa,\epsilon)$-parameter 
values. This is because of a natural trade-off between efficiency and extracted power which has triggered a number 
of recent studies to identify conditions where this trade-off is optimized \cite{BS20,CA75,SST16,W14,BSC11,BSS13,
AL20}. Our proposal shed new light in this direction since it identifies as an optimal domain for the design of cycles 
${\cal C}$, the parameter space in the proximity of an EP.

\section{Conclusions}

We have theoretically proposed and numerically demonstrated, a dramatic enhancement of the performance of thermal 
motors when they are operating in a parametric domain which is in the proximity of an EP degeneracy. The latter appears 
in the spectrum of the effective non-Hermitian Hamiltonian that describes the open circuit and it is achieved via a judicious 
(differential) coupling of the isolated circuit with the ambient baths. In the proximity of the EP, the eigenvector basis collapse 
(eigenvector degeneracy), leading to an enhanced spectral work density ${\cal W}(\omega)$. In typical circumstances, 
${\cal W}(\omega)$ is anti-symmetric with respect to the position of $\omega_{EP}$ leading to a near-zero total work $W$. 
When, however, the spectral work density ${\cal W}(\omega)$ is de-symmetrized, the total extracted power and the motor 
efficiency can acquire their maximum values in the domain of the parameter space which is in the vicinity of the EP. We 
have shown that this de-symmetrization can occur either via an explicit ${\cal PT}$-symmetry violation of the unperturbed 
system or via a spontaneous symmetry where, however, one needs to supplement it with additional spectral filtering of 
the radiation of the bath. 

Our results pave the way towards the development of a new generation of optimal thermal motors that utilize engineered 
non-Hermitian spectral degeneracies. The proposed scheme can find applications for on-chip photonics (e.g. self-powered 
micro-robots or micro-pumps in microfluidics), and electromechanical systems for harvesting ambient noise for powering a 
variety of auxiliary systems. It will be interesting to extend our study of motor efficiency to cases where the closed path in 
the parameter space is in the proximity of an EP degeneracy of higher order. It is plausible that the higher-order divergence 
of the resolvent will lead to a further enhancement of the total work. Similar questions emerge in the case where there are 
more than one EPs in the proximity of the closed path in the parameter space. The possibility to extend these design 
schemes for the realization of optimal quantum motors \cite{BC19} is also another promising direction. These, and 
other, questions will be addressed in a separate publication.


\newpage 
\onecolumngrid

\begin{center}
\textbf{\large Supplementary Material: ``Thermal Motors with Enhanced Performance due to Engineered Exceptional Points''}
\end{center}

\setcounter{equation}{0}
\setcounter{figure}{0}
\setcounter{table}{0}
\setcounter{page}{1}
\makeatletter
\renewcommand{\theequation}{S\arabic{equation}}
\renewcommand{\thefigure}{S\arabic{figure}}
\renewcommand{\bibnumfmt}[1]{[S#1]}
\renewcommand{\citenumfont}[1]{S#1}

\begin{center}
\textbf{\large SI. Expressions for forces and work}
\end{center}

In this section we derive the expressions for the forces and work in terms of the instantaneous 
scattering matrix of the associated photonic network. In particular, we arrive to Eq. (5) of the main text.

The starting point is the definition of the force, Eq. (2) 
of the main text, which, in turn, requires the knowledge of the field amplitude.
The later is given by the Coupled-Mode Theory (CMT) in Eq. (3). 
Here, we assume that the dynamical time scales of the 
mechanical degree of freedoms (MDFs) $\vec{x}$ are much slower than the photonic time-scales,
i.e., we invoke the Born-Oppenheimer approximation. 
Under this approximation, we have the CMT in frequency domain (we use the convention $f(t)=\int_{0}^{\infty} f(\omega) e^{-\im \omega t}d \omega$ for the Fourier transform)
\begin{eqnarray}
\Psi(\omega) &=& \im G^{\vec x}(\omega) D^T \theta^{(+)} (\omega); \ \ G^{\vec x}=(\omega I_N - H_{\rm eff.})^{-1}; \nonumber \\
\theta^{(-)} (\omega)&=& S^{\vec x}(\omega) \theta^{(+)}(\omega); \ \ S^{\vec x} (\omega)= -I_{N_\alpha} + \im D G^{\vec x}(\omega) D^T,
\label{G-S}
 \end{eqnarray}
where $I_m$ is the $m\times m$ identity matrix and we assume that $D$ and independent of  $\omega$ and ${\vec x}$. 
Here, the field amplitude $\Psi$ and the outgoing scattering field $\theta^{(-)}$ are dictated by the
``frozen'' or ``instantaneous" effective Hamiltonian $H_{\rm eff}=H_0({\vec x}) -\im \frac{D^T D}{2}$,  the Green function $G^{\vec x} (\omega)$, and the Scattering matrix $S^{\vec x}(\omega)$. 
In order to keep the notation simple, from now on we will drop the index ``${\vec x}$" and the dependence on ${\vec x}$ of $G^{\vec x} \equiv G$ and $S^{\vec x}\equiv S$ will remain implicit.

We get the generalized force by starting from Eq. (2) 
and using Eqs. (\ref{G-S}) and (4) 
\begin{eqnarray}
{\vec F}_{\rm av} &=& -\hbar \int_0 ^\infty \frac{d\omega}{2\pi} \sum_\alpha {\tilde \Theta}_\alpha (\omega) 
\left[ D G^\dagger \left( \nabla_{\vec x} H_0  \right) G D^T\right]_{\alpha, \alpha} \notag \\
&=&  \int_0 ^\infty \frac{d\omega}{2\pi} \sum_\alpha {\tilde \Theta}_\alpha (\omega) 
\left( \frac{\hbar}{\im} S^\dagger \nabla_{\vec x} S  \right)_{\alpha, \alpha},
\label{FScatt}
\end{eqnarray}
where $\alpha$ labels both, the reservoir and the resonator coupled to it.
To arrive to the second line we used that \cite{BKVEO11Sup}
\begin{equation}
 S^\dagger \nabla_{\vec x} S = -\im D G^\dagger \nabla_{\vec x} H_0 G D^T,
\end{equation}
which can be probed by using Eq. \ref{G-S} and by noticing that
$( G)^{-1} - (G^\dagger)^{-1} = \im D^T D$ and $\nabla_{\vec x} G = G \nabla_{\vec x} H_0 G$.

These equations are adequate to predict the forces and hence, energy extraction
capabilities, benefiting from different experimental situations.
While the first equation requires the energy distribution inside the resonators of the photonic circuit 
via the Green's function, the second equation utilizes the scattering coefficients of the circuit.

Finally, we calculate the energy extraction capability of our motor,
$W = \oint_C {\vec F}_{\rm av} \cdot d{\vec x}$,
by integrating the force when the generalized coordinates ${\vec x}$ move along a path $C$,
resulting in Eq. (5) 
of the main text.

\begin{center}
\textbf{\large SII. Work density in terms of the Green's functions}
\end{center}
In this section we derive the analytical expression for the work density, Eq. (7) in terms of the Green's functions. 
For simplicity, from now on we will consider that only diagonal elements of the Hamiltonian change with ${\vec x}$, i.e.
$\left(\nabla_{\vec x} H_0 \right)_{n,m} = (\nabla_{\vec x}  H_0 )_{n,n} \delta_{n,m} $.
Next, we consider a driving protocol such that in our photonic network only two resonators are driven, 
which we denote with indexes $n=p,q$; and 
each one of the resonant frequencies of those resonators $\omega_n=(H_0)_{n,n}$
depend on only one coordinate $x_\nu$, 
i.e. $\frac{\partial \omega_n}{\partial x_\nu}=\frac{\partial \omega_n}{\partial x_n}\delta_{n,\nu}$. By using Eq. \ref{FScatt}, we calculate the work as $W = \oint_C {\vec F}_{\rm av} \cdot d{\vec x}=
\iint_A (\nabla_{\vec x} \times {\vec F}_{\rm av}) \cdot d{\vec A}$, i.e.,
\begin{eqnarray}
 W &=&  -\hbar \int_0^{\infty} \frac{d\omega}{2\pi} \sum_\alpha 2\gamma_\alpha {\tilde \Theta}_\alpha (\omega) 
  \oint_C \left( |G_{p,\alpha}|^2 \frac{\partial \omega_p}{\partial x_p} dx_p + 
 |G_{q,\alpha}|^2 \frac{\partial \omega_q}{\partial x_q} dx_q\right), \nonumber \\
   &=& -\hbar \int_0^{\infty} \frac{d\omega}{2\pi} \sum_\alpha 2\gamma_\alpha {\tilde \Theta}_\alpha (\omega) 
    \iint_A \left( \frac{\partial |G_{q,\alpha}|^2}{\partial x_p} \frac{\partial \omega_q}{\partial x_q} 
    - \frac{\partial |G_{p,\alpha}|^2}{\partial x_q} \frac{\partial \omega_p}{\partial x_p}\right) dx_p dx_q,
       \label{eq:work2}
\end{eqnarray}
where, to arrive to the second equality, we have used Green's theorem.
It is useful to turn Eq. \ref{eq:work2} into a more compact expression by using the geometric 
integral $P_\alpha$, defined in Eq. (5).
Finally, the work density per unit area
\begin{equation}
 {\cal W}_\alpha = \lim _{A\rightarrow 0} \frac{P_\alpha}{A} = 4 \gamma_\alpha \hbar{\rm Re} \left( G_{p\alpha}^* G_{pq} G_{q\alpha} 
    - G_{q\alpha}^* G_{qp} G_{p\alpha} \right),
    \label{eq:work_density_Green}
\end{equation}
where, we have used $ \frac{\partial |G_{nm}|^2}{\partial x_j} = 2 {\rm Re} \left( G^*_{nm} G_{nj}\frac{\partial \omega_j}{\partial x_j} G_{jm}\right),$
and since $A= \iint_A \frac{\partial \omega_p}{\partial x_p}\frac{\partial \omega_q}{\partial x_q} d x_p d x_q\rightarrow0$, the Green's functions are evaluated at the center of the loop $\{x_p,x_q\}$.

\begin{center}
\textbf{\large SIII. Green's function near the EP}
\end{center}
When the path $C$ is performed in the vicinity of an EP, the interplay of the coalescing resonances can 
dramatically affect the response of the system under small perturbations.
This abrupt behavior can be related to the Green's function, which in the vicinity of a (second order) EP,
presents a sharp Lorentzian-squared resonance, evidenced through the modal expansion around the EP\cite{PZMHHRSJ17Sup}
\begin{equation}
 G(\omega)=\sum_n \frac{1}{\omega-\omega_n} \frac{\tilde{u}_n \cdot \tilde{v}_n^T}{\tilde{v}_n^T \cdot \tilde{u}_n}
 \approx \frac{A}{\omega-\omega_{EP}}  + \frac{B}{(\omega-\omega_{EP})^2} .
 \label{Green_modal_exp}
\end{equation}
To arrive to the right hand side, we assume $\omega\approx \omega_{EP}$ and then we can restrict the summation to $n$ indexes whose eigenfrequencies $\omega_n$, and the corresponding right (left) eigenvectors $\tilde{u}_n$ ($\tilde{v}_n$) of $H_{\rm eff}$, are associated to the EP. To simplify the notation, 
we denote them by $\omega_\pm$ and $\tilde{u}_\pm$ ($\tilde{v}_\pm$). 
Next, we use an expansion in a Newton-Puiseux series
invoking fractional-powers of a perturbation parameter $p \ll 1$ \cite{SM03Sup}
\begin{eqnarray}
 \omega_\pm &=& \omega_{EP} \pm p^{1/2} \lambda_1 + p \lambda_2 \pm p^{3/2} \lambda_3 + \ldots ,\nonumber \\
 \tilde{u}_\pm &=& u_0 \pm p^{1/2} \lambda_1 u_1 + p w_2 \pm p^{3/2} w_3 + \ldots ,
 \label{Puiseux}
\end{eqnarray}
where a similar equation holds for $\tilde{v}_\pm$ and the Hamiltonian $H_{\rm eff}=H_{\rm eff}^T\approx H_0 + p H_1 + \cdots$. The EP occurs at $p=0$, and there, the defective right (left) eigenvector $u_0$ ($v_0$) of $H_0$ and the associated Jordan vector $u_1$ ($v_1$) satisfy the Jordan chain relations
\begin{eqnarray}
 H_0 u_0 &=& \omega_{EP} u_0 \ ; \ H_0 u_1 = \omega_{EP} u_1 + u_0 \nonumber \\
 v_0 ^T H_0 &=& \omega_{EP} v_0^T \ ; \ v_1^T H_0 = \omega_{EP} v_1^T + v_0^T ,
\end{eqnarray}
with the normalization conditions $v_0 ^T u_1 = 1$ and  $v_1^T u_1 =0$ and the properties 
$v_0^T u_0=0$ and $v_1^T u_0 = v_0 ^T u_1 =1 $. It follows that $\lambda_1=\pm \sqrt{v_0^T H_1 u_0}$, 
which determines the leading order of the expansions in Eqs. \ref{Puiseux}.
By keeping these expansions up to leading order, we arrive to 
the right hand side of Eq. \ref{Green_modal_exp}, where $A=u_1 v_0^T + u_0 v_1^T$ and $B=u_0 v_0^T$.



\begin{thebibliography}{1}
\bibitem{CMTpar} The CMT Hamiltonian's parameters that best fit scattering spectrums of the circuit setup are 
$\gamma_1\approx3.6\times10^{-3}$, $\gamma_2\approx3.2\times10^{-4}$, $\omega_1\approx 1-\kappa$, and $\omega_2\approx 0.94-\kappa$,
in units of $\omega_0=2\pi\times 1MHz$.
For a detailed model of the CMT see \cite{FLK20}
\end{thebibliography}
\end{document}